%% file: main.tex
\documentclass[11pt]{article}

\usepackage[T1]{fontenc}
\usepackage{amsmath}
\usepackage[french]{babel}

\usepackage[top=2.0cm, bottom=2.0cm, left=2.5cm, right=2.5cm, headheight=0cm]{geometry}

\RequirePackage{relsize}
\RequirePackage[dvipsnames,svgnames]{xcolor}
\RequirePackage{tikz}
\usetikzlibrary{shapes,shadows}

\usepackage{url}
\usepackage{enumitem, txfonts}

\usepackage{eso-pic} 
\usepackage{fontawesome}

\usepackage{enumitem}
\usepackage{amssymb}
\usepackage{marvosym}

\usepackage{subcaption}

\usepackage[most]{tcolorbox}

\usepackage{ragged2e}

\usepackage{graphicx}
\usepackage{caption}
\usepackage{lipsum}

\usepackage{hyperref}
\usepackage{url}

\usepackage{pythonhighlight}

\usepackage{pdfpages}

\usepackage{xcolor,colortbl}
\usepackage{array}
\usepackage{tabularx}

\usepackage{multirow}
\newcolumntype{M}[1]{>{\centering\arraybackslash}m{#1}}
\newcolumntype{L}[1]{>{\raggedright\let\newline\\\arraybackslash\hspace{0pt}}m{#1}}
\newcolumntype{C}[1]{>{\centering\let\newline\\\arraybackslash\hspace{0pt}}m{#1}}
\newcolumntype{R}[1]{>{\raggedleft\let\newline\\\arraybackslash\hspace{0pt}}m{#1}}

\usepackage{colortbl}

\definecolor{darkgreen}{rgb}{0.0, 0.5, 0.0}

\newcommand{\blap}[1]{\vbox to 0pt{#1\vss}}
\newcommand\AtUpperLeftCorner[3]{%
	\put(\LenToUnit{#1},\LenToUnit{\dimexpr\paperheight-#2}){\blap{#3}}%
}
\newcommand\AtUpperRightCorner[3]{%
	\put(\LenToUnit{\dimexpr\paperwidth-#1},\LenToUnit{\dimexpr\paperheight-#2}){\blap{\llap{#3}}}%
}

\makeatletter

\tcbset{
	colframe=blue!50!black,  
	colback=blue!5!white,    
	boxrule=1mm,             
	sharp corners,           
	width=\textwidth,        
}

\begin{document}
	
	\title{An extensive analysis and calibration of the Modular Aggregation Algorithm across three categories of for GNSS trajectories data sources}
	\author{%
		Marie-Dominique Van Damme*, Yann M\'eneroux, Ana-Maria Olteanu-Raimond \\
		Univ Gustave Eiffel, IGN-ENSG, LASTIG\\
		73 Avenue de Paris \\
		Saint-Mand\'e \\
		France 
	}
	\date{\today}

%
%

\input{1_couverture.tex}
    \input{2_intro}

\input{3_data}				%
    \input{4_algo_LE}
    \input{4_MIAA}

    \input{5_extension}

	\newpage

    \bibliographystyle{alphaurl}
	\bibliography{biblio.bib}

\end{document}

%% file: 1_couverture.tex
\begin{titlepage}
    \enlargethispage{2cm}
 
    \AddToShipoutPicture{
        \AtUpperLeftCorner{1.5cm}{2.2cm}{\includegraphics[width=8cm]{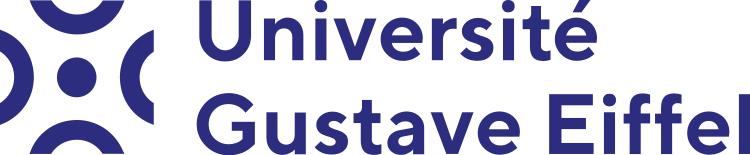}}
        \AtUpperRightCorner{1.5cm}{1.6cm}{\includegraphics[width=4.0cm]{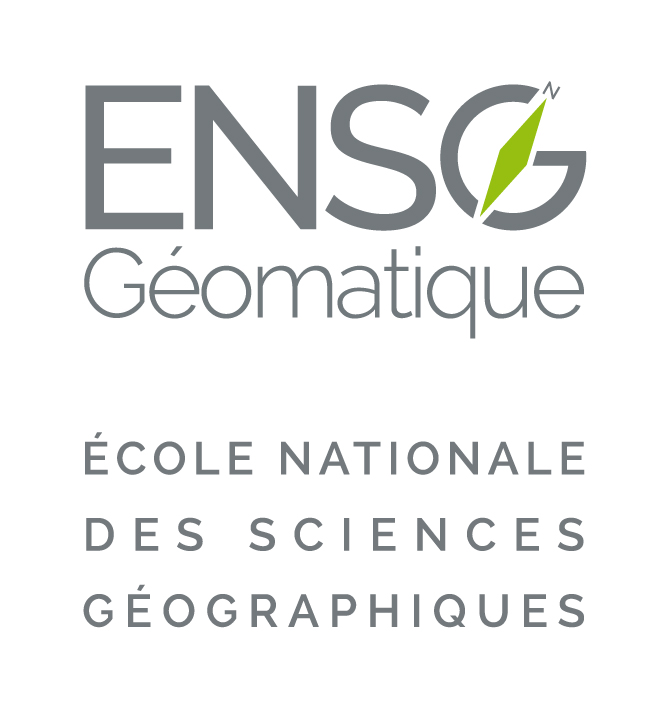}}
        \AtUpperRightCorner{5.5cm}{1.6cm}{\includegraphics[width=4.0cm]{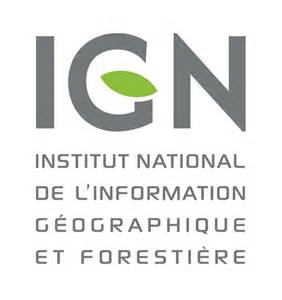}}
        
       \AtUpperLeftCorner{1.5cm}{4.2cm}{\includegraphics[width=8cm]{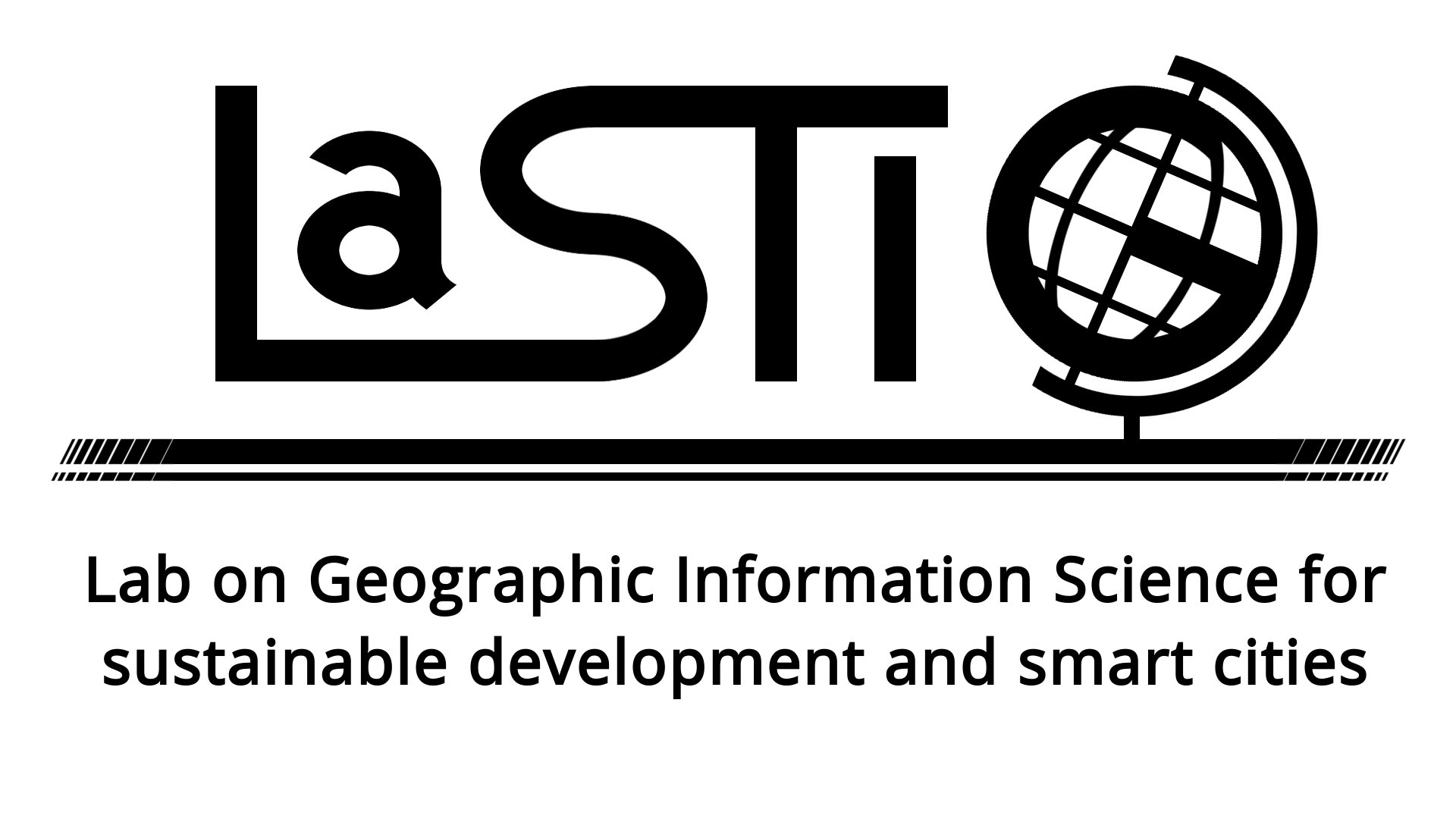}}

         \AtUpperRightCorner{5.5cm}{6.2cm}{\includegraphics[width=4.0cm]{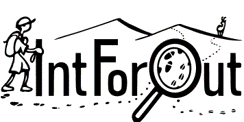}}        

    }

\vspace*{8cm}

\begin{tcolorbox}
	\centering
	\maketitle
	
\flushleft
\vspace*{0.5cm}
\small{* All authors contributed equally to this research.} \\
version 2.0
\end{tcolorbox}
 
%
%

\begin{center}
\end{center}

%
%
%
 
    
    \vspace{1.5cm}
    
    
    \vspace{4.5cm}

\end{titlepage}
\ClearShipoutPicture

%% file: 2_intro.tex
\tableofcontents
\newpage

\listoffigures
\newpage

\section{Introduction}

\noindent With the growing interest in outdoor recreation, there has been a surge in sport associations proposing new group outdoor activities and offering new opportunities for people to explore and engage with nature in exciting ways. This expansion contributes to the economic development of local communities and, more generally, to tourism. However, the expansion of outdoor activities could have negative consequences on the preservation of biodiversity.
Indeed, scientific ecologists mentioned "the landscape of fear" to describe the negative effects on wildlife (e.g., chamois) disturbance \cite{gruas2022} or loss of plant species diversity.

\vspace{0.5\baselineskip}

\noindent In this context, it becomes crucial to better estimate the human pressure in time and space to guide informed management
and conservation actions as well as to compute SDGs indicators.
Such data measuring the presence of people, animals, and their interactions or protected plants are collected by local and regional stakeholders. However, a previous report showed a lack of official data and disparities between countries at global scale. In parallel, recent research has demostrated the potential of crowdsourcing and citizen science data to enhance the representation of human pressure on ecosystems such as GNSS trajectories collected and shared openly by
practitioners on different platforms (e.g. CamptoCamp, IGNRando,
OutdoorVision, OpenStreetMap, Geo-life) as well
as data provided by citizens through Citizen Science (CS) initiatives.

\vspace{0.5\baselineskip}

\noindent In this context, our aim is to develop innovative methods and tools to assess human pressures on ecosystems and provide new solutions to help stakeholders mitigate these impacts while promoting sustainable tourism and outdoor activities. More concretely, the goal is to build a framework allowing integration of spatiotemporal data representing a network of outdoor activities derived from road network and GNSS trajectories. We aim to enhance the semantic of this network with human frequency and activities types, biodiversity information (e.g. rest areas for animals) as well as human-wildlife encounters. One of the challenges is to deal with data heterogeneity originating from multiple sources across different spatial and temporal scales and from various types of data producers, including national, regiona, and local authorities, research communities, and citizens (through participatory science or volunteered geographic information). Both input and output data, as well as the proposed methods, will be described in a knowledge graph to facilitate data retrieval, dissemination, and reproducibility.   

\vspace{0.5\baselineskip}

\noindent This work is part of the research project IntForOut (Multisource spatial data INTegration FOR the Monitoring of Ecosystems under the pressure of OUTdoor recreation) funded by the French National Research Agency (n° ANR-23-CE55-0003).

\vspace{0.5\baselineskip}

\noindent This technical report aims to complement the conference paper \cite{10.1145/3678717.3691325} by providing additional experiments or further details that could not be included in the paper.

\vspace{1\baselineskip}

\subsection{Data availability}

\begin{itemize}
	\item This extensive analysis is based on the work described in \cite{10.1145/3678717.3691325} ;
	\item The open-source implementation of the MIAA algorithm in the tracklib library \cite{menerouxtracklibv2} ;
	\item Multi-sensors and multi-canopy traces acquisition with the ground truth \cite{MOUSLS2025}
\end{itemize}

%% file: 3_data.tex
\newpage
\section{Data and material} 

\noindent Since our goal is to reconstruct the geometry of the path from a given subset of GPS trajectories, disregarding the time dimension, each dataset used in this study contains trajectories following the same route, travelling in the same direction, and having approximately the same start and end points. In this study, the data used consist of a collection of datasets that respect these conditions, aligning with those outlined in \cite{etienne:hal-02110086}. 

\vspace{0.5\baselineskip}

\noindent The scenarios of greatest interest for studying the aggregation method are paths under forest cover, due to the noise effect, and those containing a series of sharp handling challenges in aggregation. Moreover, our analysis will be carried out by using both synthetic and real trajectories. Thus three categories of data are considered: 
\begin{enumerate}
	\item Realistic Synthetic GNSS Trajectories: These simulated trajectories provide a controlled environment to test the algorithm's performance under various predefined conditions, allowing us to assess its accuracy and robustness; 
	
	\item Multi-Sensor Trajectories: data collected with different sensors (professional, watched and smartphone applications) to evaluate how well the algorithm handles the heterogeneity of sensor data. Multi-sensors trajectories are acquired by following a repeatable data collection protocol, we defined;  
	
	\item Real world data (i.e. crowdsourced datasets), more specifically three crowdsourced datasets categories (terrain-constrained itinerary, infrastructure-constrained and multi-scale); this choice is not strictly «metrological» but rather qualitative: to investigate graphically how the algorithm performs on real typical cases.
\end{enumerate}

\subsection{Synthetic GNSS trajectories}

\noindent In order to assess the metrological performances of the algorithm (\textit{i.e.} its ability to reconstruct accurately the common path followed by all the individual sample trajectories) it is required to test it on a large array of configurations, with a substantial amount of GPS trajectories. This raises two operational problems: 

\vspace{0.25\baselineskip}

\begin{itemize}[label=$\thicksim$]
\item First, to compare the estimated trajectory with the real route actually followed by the individual samples, it is required to perform a costly and time-consuming on-the-field survey to measure the ground truth. This is especially problematic since moderately to densely covered forest areas are of interest, where the direct use of precision carrier-phase GNSS with Real-Time Kinematic fast and convenient procedure is often impossible. Therefore, ground truth has to be measured with classical surveying traverse, tied to an absolute reference point, possibly located hundreds of meters away; spending a full day of topometric survey for each case study is not realistic. 

\vspace{0.25\baselineskip}

\item Secondly, for each case study, individual GNSS trajectories must be collected by walking several times along the ground truth route. Because GNSS error is known to be auto-correlated in time \cite{roberts1993gps}, for a realistic acquisition of data, trajectories must ideally be sampled at different times of day, which increases again drastically the effort needed to collect real data. 

\end{itemize}

\vspace{0.5\baselineskip}

\noindent To overcome these limitations, it was first decided to proceed to extensive experimentation of the algorithms on simulated GNSS trajectories. This enables to simulate as many case studies as needed, with potentially unlimited number of GNSS trajectories, and a readily available  ground truth track to compare the results with. 

\begin{figure}[ht!]
	\centering
	\includegraphics[width=5cm]{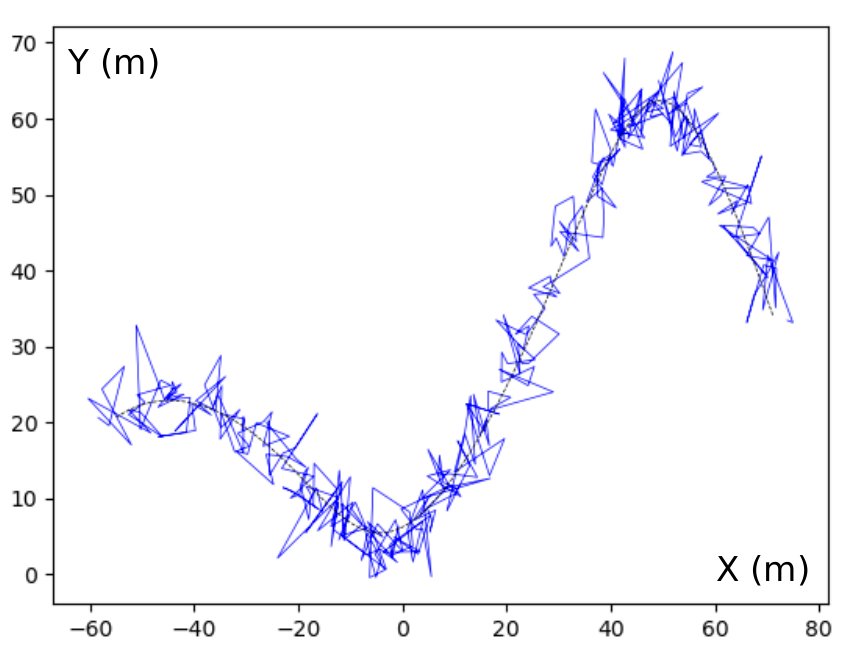}
	\caption{Example of a non-realistic simulation of a GNSS trajectory with a white noise process, completely missing out the true correlation pattern of GNSS measurements.}
	\label{simu1}
\end{figure}

\vspace{1\baselineskip}

\noindent This methodology however, requires an accurate modeling of auto-correlation error of GNSS trajectories, to avoid non-realistic simulations, as depicted for example in Fig. \ref{simu1}, which can also result in topological errors \cite{bonin2002modele, vauglin1997modeles}. 

\vspace{1\baselineskip}

\noindent GNSS errors were then modeled through their covariance function

\hspace*{2em} $\gamma(s_1, s_2) = \mbox{Cov}(X(s_1), X(s_2))$ 

\noindent describing the statistical covariance between positioning errors $X(s_1)$ and $X(s_2)$ at two locations $s_1$, $s_2 \in \mathbb{R}_+$ (described trough their curvilinear abscissa along the ground truth trajectory). 

\vspace{0.5\baselineskip}

\begin{figure}[ht!]
	\centering
	\includegraphics[width=8cm]{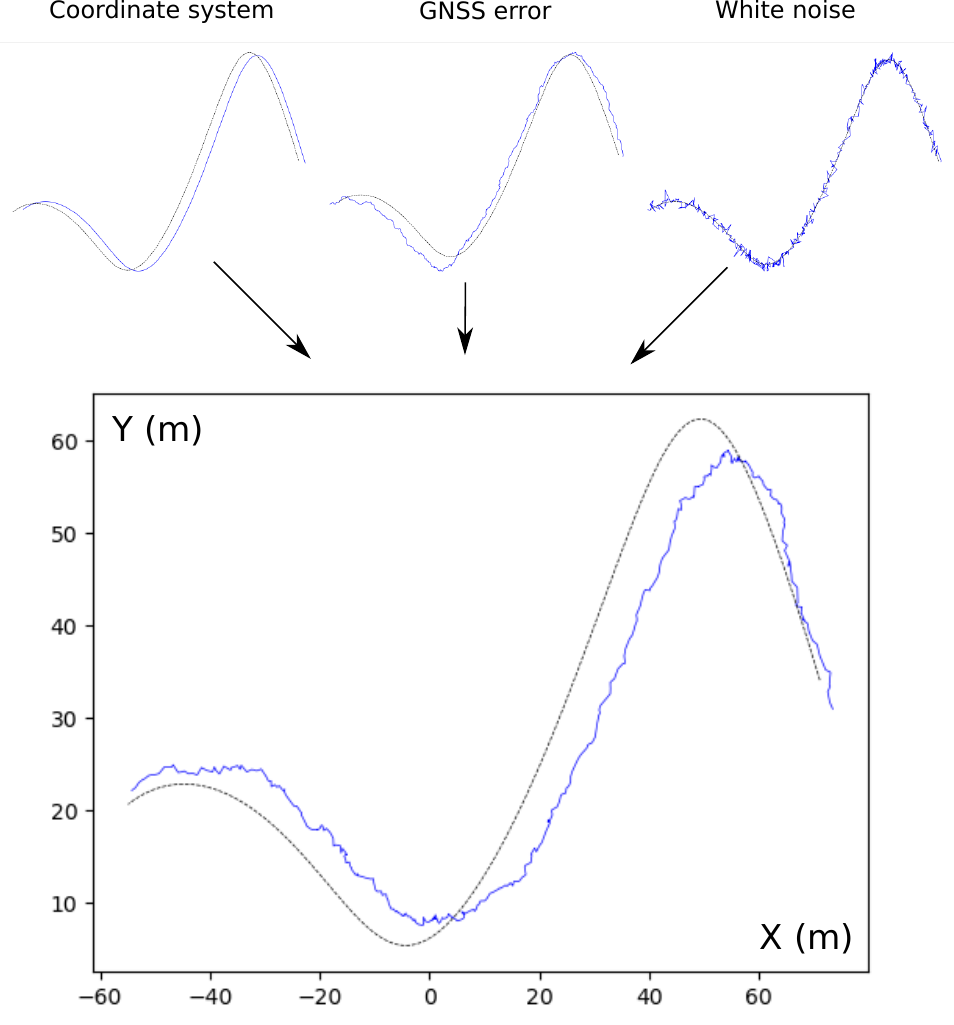}
	\caption{Illustration of the three error components in the GNSS trajectories. From left to right: (1) a long wave-length process describing coordinate system errors, (2) an intermediate wave-length process describing GNSS observation errors (auto-correlated in space and time) and (3) a white noise process (e.g. heat, vibrations, electronic noise). }
	\label{simu2}
\end{figure}

\noindent Further, the error $X$ is supposed to be a second-order stationnary process (hence described only by the difference $s_2-s_1$), and is modelled to take into account different error components in the GNSS trajectory measurement process, as illustrated on Fig \ref{simu2}. 

\begin{figure}[ht!]
	\centering
	\includegraphics[width=6.5cm]{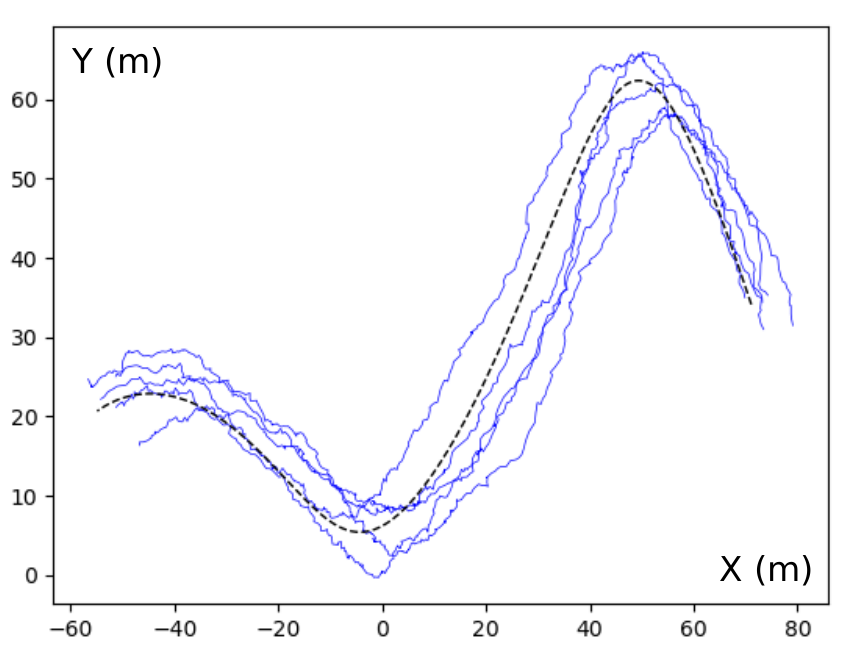}
	\caption{Example of 5 synthetic GNSS trajectories (blue) generated on a common ground truth track (dashed line).}
	\label{simu3}
\end{figure}

\noindent Generation of GNSS trajectories was done independently on each of the two planimetric components, with a methodology described in \cite{ripley2009stochastic} and also employed in \cite{Meneroux2023}: with a random generator, we sampled $n$ i.i.d. unit-variance and zero-mean gaussian values, compiled in a vector $\mathbf{x}$. It can easily be shown that, for any positive-definite matrix
$\mathbf{\Sigma} \in \mathbb{R}^{n \times n}$, the random vector $\mathbf{y} = \mathbf{Ax}$ where $\mathbf{A}$ is a Cholesky factor of $\mathbf{\Sigma}$, is a realization of a correlated random vector $\mathbf{Y}$ having covariance matrix $\Sigma$. The covariance matrix $\mathbf{\Sigma}$ is formed with $\mathbf{\Sigma}_{ij} = \gamma(s_j-s_i)$ (\textit{Cf.} example in Figure \ref{simu3}).

\noindent \underline{\textbf{Application:}} In \cite{10.1145/3678717.3691325}, the used methodology is the following: for each case study, a reference track is simulated (or extracted from an existing topographic database) and is considered as the ground truth track from which all GNSS trajectories are simulated. The error between the estimated and ground truth track is then evaluated, which in turn, enables to assess the sensitivity of the algorithm to all its parameters. Similar methodologies have been used for example by \cite{biljecki2015propagation} and \cite{zhang2015cross}.

\vspace{0.5\baselineskip}

\noindent In our experimentation, trajectories have been generated with a noise generated with a 100 m range Gaussian Process and with a 50 cm-amplitude for referencement error, completed by a noise generated with a 5 m-amplitude exponential covariance process (GPS error) (\cite{grejner2005improving}) with a 50 m correlation scope and completed by 1 m white noise process (vibrations, electronic noise, etc.). \\

\noindent Which translates to the following code using the \textit{Tracklib} library (\cite{menerouxtracklibv2}):

\vspace{0.5\baselineskip}

\begin{python}
# Generate a track
tkl.seed(123)
base_lacets = tkl.generate(0.4, dt=10)
chemin = tkl.noise(base_lacets, 20, tkl.SincKernel(20),
                    direction=tkl.MODE_DIRECTION_ORTHO)[::3]
chemin = chemin[80:250]
chemin.scale(10)
	
# Generate a collection of five identical tracks
N = 5
tracks3 = tkl.core.TrackCollection([chemin]*N)

# Generate noise to get realistic GNSS trajectory
tracks3.noise(0.5, tkl.GaussianKernel(100))
tracks3.noise(5, tkl.ExponentialKernel(50))
tracks3.noise(1, tkl.DiracKernel())

# Visualisation
plt.plot(chemin.getX(), chemin.getY(), color="black", linestyle='--', linewidth=1)
for track in tracks3:
    plt.plot(track.getX(), track.getY(), color="royalblue", linestyle='-', linewidth=1)
\end{python}

\newpage
\subsection{Multi-sensors and multi-canopy traces acquisition}

\noindent To asses the impact of canopy and the sensors on precision accuracy, we defined and implemented a data collection protocol. \\

\noindent First, based on the literature, we identified three types of canopy (i.e. open area, moderate coverage, and heavy coverage). Additionally, we delineated five types of sensors (i.e. mobile phone equipped with VisioRando application, Polar GPS device, Garmin GPS device, Keymaze device and professional Ublox GPS sensor chip). 

\vspace{0.5\baselineskip}

\noindent Second, for each type of canopy, the following data collection protocol was defined:
\begin{itemize}[label=$\thicksim$]
	\item Identification of areas without spatial constraints (e.g. bridge, stream, unobstructed)
	\item Identification within a rectangle of an Origin/Destination route with moderate winding and approximate length of 300 m. 
	\item Collect five round trips following the route exactly. 
\end{itemize}

\vspace{0.5\baselineskip}

\noindent Third, the field work was done by two of the authors of this paper. The placement of the sensors is also relevant according to the literature \cite{Blunck2011}. To limit this effect, the GPS watches were worn on the wrist, the professional GPS devices were carried in a bag with the antenna positioned externally, and the mobile phones were held in the hand. Data collection has been carried out during the summer season. In total, for the three types of canopy and five sensors, 150 trajectories are collected.

\vspace{0.5\baselineskip}

\begin{figure}[h]
	\centering
	\includegraphics[height=7.0cm]{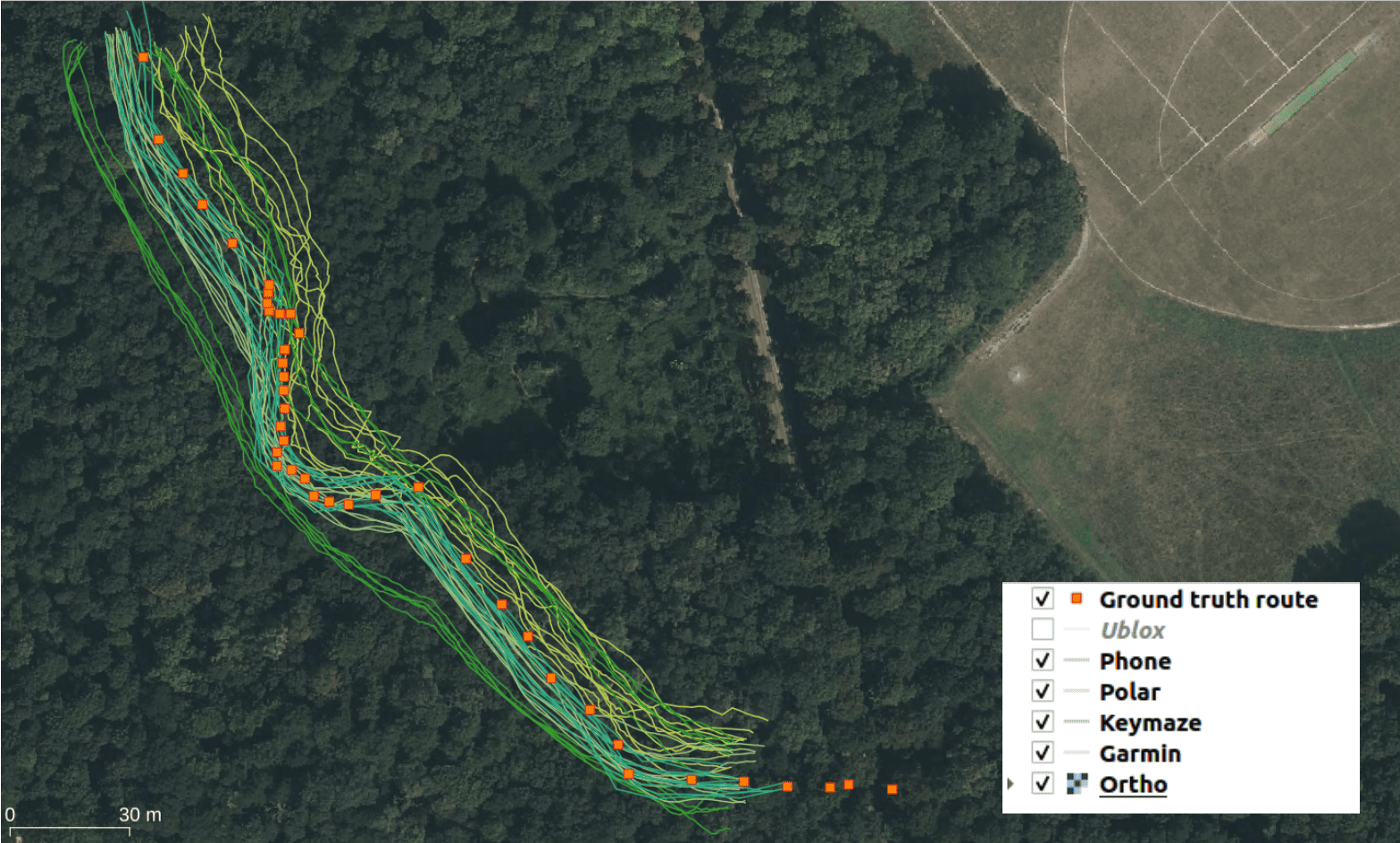}
	\caption{Set of 50 trajectories collected with five sensors in dense forest and ground truth route.}
	\label{fc}
\end{figure}

\noindent \underline{\textbf{Application:}} In \cite{10.1145/3678717.3691325}, our experiments focus only on dense forest 50 trajectories collected with five sensors are shown in Figure \ref{fc}.

\vspace{1\baselineskip}

\noindent The ground truth route is obtained through a topometric survey conducted by a group of students as part of a topometric project, under the supervision of the authors of this paper and their teachers \cite{rapport2024pir}. Absolute positioning is conducted through GNSS differential static positioning of a set of reference points located in open-sky conditions, between 300 and 500 meters from the surveyed route. Topometric determination of the route geometry is performed with surveying traverse. The output ground truth is sampled with about 42 points (\textit{i.e.} about 1 point every 7 meters) with an absolute positioning accuracy of 5 mm in each 3D axis (1$\sigma$) (\textit{Cf.} Figure \ref{fc}).

\subsection{Crowdsourced traces}

\noindent To validate the results obtained with synthetic GNSS tracks and data acquired according to the proposed protocol, we have chosen to test the algorithms on crowdsourced trajectories, as these will be used to derive pressure  and route frequency indicators for the end-users. 

\vspace{0.5\baselineskip}

\noindent To this end, our third experiment was carried out considering tracks downloaded from Visorando\footnote{\url{www.visorando.com}} and from Wikiloc \footnote{\url{www.wikiloc.com}}, websites offering donwloading tracks published online by contributors. The GPS sensors used are therefore unknown, as well as data changes, context conditions. 

\vspace{0.5\baselineskip}

\begin{figure}[h]
	\centering
	\includegraphics[width=15.0cm]{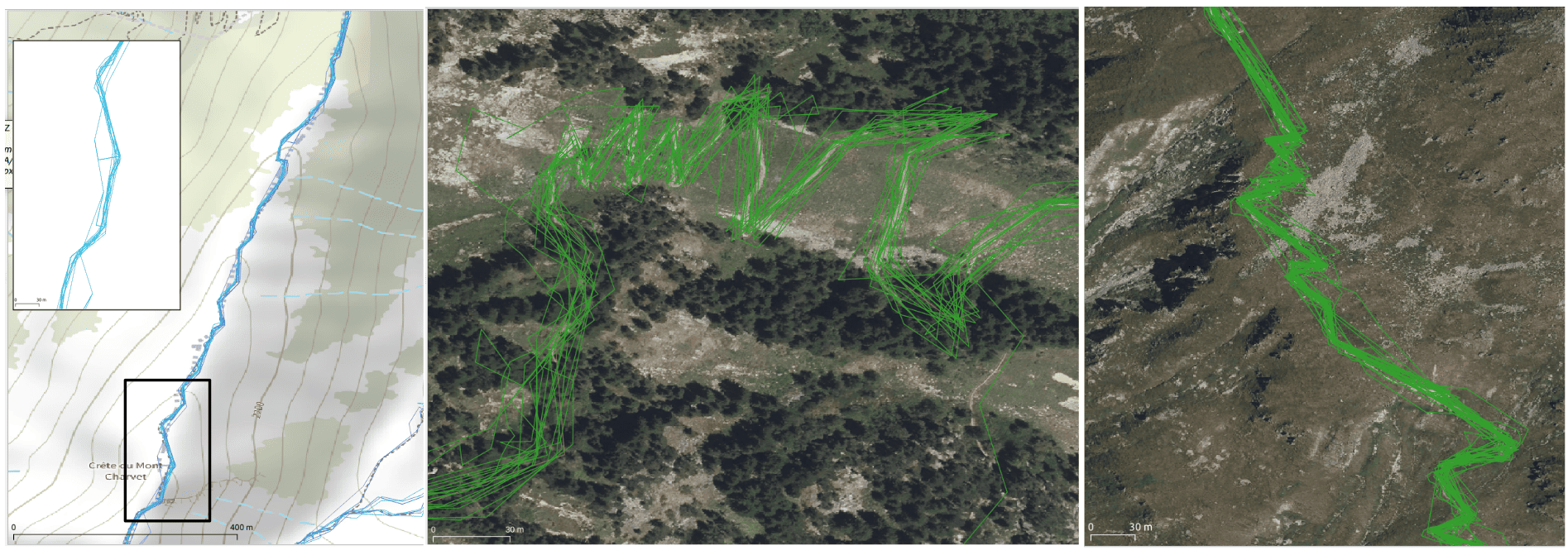}
	\caption{Trajectories in three differents contexts with spatial constraint: (a) ridge (terrain constraint), (b) series of sharp (infrastructure constraint), (c) heterogeneous trail shape (scale variation)}
	\label{3zones3constraints}
\end{figure}

\noindent In this experiment, we focused on hiking activity in the mountain area of the Pralognan Valley in the French Alps. Specifically, from all the trajectories, we selected some of them having specific and challenging configurations and spatial constraints: terrain constraints (e.g. ridge, river), infrastructure support (\textit{e.g.} switchback trail) and variations of the shape of the path (\textit{e.g.} a straight line followed by a series of twists and turns with varying distances between them). Note that, in a dataset, all the selected trajectories have the same origin/destination (\textit{Cf.} Figure \ref{3zones3constraints}). Note that, traces may have been previously filtered via a simplification algorithm like Douglas-Peucker. Indeed, all geometries offer the same characteristics like vertex of bends, turns in route. If this was not the case, traces would be more dissimilar in shape.


\noindent As the reviewers suggested in \cite{10.1145/3678717.3691325}, considering a broader approach based on real world data would be interesting. This could be done by generalizing the first context: a route with terrain constraints is a corridor, covered by an acceptable number of traces, for which we can guarantee they have actually followed the path. For example, in addition to ridge paths, constrained route segments can include trails along the mountainside, trail sections bordered by fences or rock walls. This can be addressed by defining a protocol to extract such trajectories, involving running spatial and topological queries to identify such areas, and then cross-referencing them with traces dataset. To ensure the representativeness of such constraint contexts, we use crowdsourced data issued from collaborative platforms. Eventually, the geometry of the road section stemming from an accurate topographic database is used as the ground truth. Alternatively, ground truth dataset may be collected directly on the field with topometric surveying. The algorithm will be run on all extracted constrained route segments, providing a quasi-metrological validation on real data. Although this approach is relevant, we could not further develop it due to the paper length constraints, so it will be explored in future works.

%% file: 4_algo_LE.tex
%
\newpage
\section{Formalization of the original algorithm}
\label{LEAlgo}

\noindent This section formalizes the algorithm proposed \cite{etienne:hal-02110086} lacking in a mathematical description to better describe its properties.

\vspace{0.5\baselineskip}

\noindent Considering a set of trajectories, where each trajectory is defined as an $n$ ordered points: \\
\hspace*{4em} $\mathcal{X} = (\mathbf{x}_i)_{i=1..n}$ , with $\mathbf{x}_i \in \mathbb{R}^2$ and GPS records in a 2D space.

\vspace{0.5\baselineskip}

\noindent Let $d$ be a distance (Euclidian, Manhattan, etc.) between those points: \\
\hspace*{4em} $d : \mathbb{R}^2 \times \mathbb{R}^2 \rightarrow \mathbb{R}_+$. \\
\noindent Note that heights and timestamps for example, could also be considered with $\mathbf{x}_i,~\mathbf{y}_i \in \mathbb{R}^3$ or $\mathbb{R}^4$, and defining $d$ accordingly to measure distance between three-dimensional timestamped records.

\vspace{0.5\baselineskip}

\noindent An \textbf{aggregated trajectory}, noted $\mathcal{AX}$, is defined as the best geometric representation of a set of trajectories $\mathcal{X}$ following exactly the same route defined from an origin to a destination:  \\
\hspace*{4em} $\mathcal{AX}$ = ($\overline{xj}$), j=1..m, \\
\noindent where $\overline{xj}$ represents the aggregated points of matched points for the j-th point in the master trajectory.

\vspace{0.5\baselineskip}

\noindent We defined an \textbf{accurate aggregated trajectory}, noted $\mathcal{AAX}$, an aggregated trajectory that optimizes a quality criterion $\mathcal{Q}$, with respect to the ground truth $\mathcal{G}$ (unknown). More formally, $\mathcal{AAX}$ optimizes $E[\mathcal{Q}]$, where E is the expected value of the quantity $\mathcal{Q}$. The main challenge is to define $\mathcal{Q}$, and then to compute it in an approximate and satisfactory way, which is our focus.

\vspace{0.5\baselineskip}

\noindent One might say, a trace $\mathbf{T}_1$ is a good partial representation of $\mathbf{T}_2$ if $\mathbf{Q}_{T2 \rightarrow T1}$ is minimal with $\mathcal{Q}$ being the square root of the mean of the squared distances between each point in $\mathbf{T}_1$ and its closest neighbor on $\mathbf{T}_2$.

\vspace{0.5\baselineskip}

\noindent Thus, the quality of the trace $\mathcal{AAX}$ can then be evaluated from the average of the partial qualities: \\
\hspace*{4em} $\mathcal{AAX} = (\mathbf{Q}_{\mathcal{AAX} \rightarrow G} + \mathbf{Q}_{G \rightarrow \mathcal{AAX}}) \div 2$, with $\mathcal{G}$ the ground truth. 


\subsection{Algorithm initialization and termination}

\noindent As mentioned before, this algorithm uses an iterative refinement approach that improve the existing solution at each step, continuously performing multiple matches. And each iteration is composed of three steps: trajectory matching, representative selection on trajectory sections and aggregation of representative points.

\vspace{0.5\baselineskip}

\noindent But first algorithm initialization consists of choosing a first trajectory, will be called the master trajectory, $\mathcal{R} = (\mathbf{r}_j)_{j=1..m}$. This process involves defining a set of reference positions (($\mathbf{r}_j)_{j=1..m}$), which are all the vertices of the master polyline. Selecting a master trajectory, to start the algorithm, avoids having to match all the trajectories in pairs.

\vspace{0.5\baselineskip}

\noindent And finally, the iterative process stops when the difference between the aggregated trace and the master trace is below a minimum threshold.

\subsection{First step: trajectory matching}

\noindent The points of each trajectory will be matched with the master trajectory. To be more precise, a matching between the trajectory $\mathcal{X}$ and the master trajectory $\mathcal{R}$ is a bipartite graph ($\mathcal{X}$, $\mathcal{R}$, $\mu$) with no isolated vertex and where $\mu$ are the set of matching links. Note that the connectivity constraint imposes that each point in $\mathcal{X}$ is linked to at least one points in $\mathcal{R}$, and reciprocally. We denote by $\mathcal{M}(n, m)$ the set of all possible matching between two trajectories containing $n$ and $m$ points respectively.

\vspace{0.5\baselineskip}

\noindent An \textbf{ordered matching} between trajectories $\mathcal{X}$ and $\mathcal{R}$ is a matching with the additional constraint that there should be no pair of \textit{crossing} links, \textit{i.e.} that we cannot find two links $(\mathbf{x}_i, \mathbf{r}_j)$ and $(\mathbf{x}_k, \mathbf{r}_l)$ with $i > j$ and $k < l$. We denote by $\mathcal{M}^+(n, m)$ the set of all possible ordered matchings between two trajectories containing $n$ and $m$ points respectively. Note that the definition of an \textit{ordered matching} imposes that $(\mathbf{x}_1, \mathbf{r}_1)$ and $(\mathbf{x}_n, \mathbf{r}_m)$, \textit{i.e.} start and end points of trajectories $\mathcal{X}$ and $\mathcal{R}$ should be matched together respectively.

\vspace{0.5\baselineskip}

\noindent Matching algorithms based on dynamic programming approach (\textit{Dynamic Time Warping matching} (DTW), \textit{Discrete Fréchet distance}) are used to calculate ordered matching $\mu \in \mathcal{M}^+(n, m)$. The $L_p$-norm optimal dynamic time warping matching between the trajectory $\mathcal{X}$ and the master trajectory $\mathcal{R}$ is given by:

\vspace{0.5\baselineskip}
\centering
$DTW_p(\mathcal{X}, \mathcal{R}) \in \underset{\mu \in \mathcal{M}^+(|\mathcal{X}|, |\mathcal{R}|)}{\mbox{argmin}}~ \sum_{i=1}^{|\mu|} d(\mathbf{x}_{\mu_i(1)}, \mathbf{r}_{\mu_i(2)})^p$ 
\justifying

\noindent where, for any index $i \in [1,|\mu|]$, the point of index $\mu_i(1)$ of trajectory $\mathcal{X}$ is linked to the point of index $\mu_i(2)$ of $\mathcal{R}$.

\vspace{0.5\baselineskip}

\noindent Note that the discrete Fréchet distance (the method used by \cite{etienne:hal-02110086}) is similar, but not identical to DTW, since Discrete Fréchet distance seeks to minimize the maximum of distances between the two curves, whereas DTW aims at minimizing the sum of these distances. In fact, the discrete DTW distance given in the formula above degenerates to Fréchet distance when $p$ grows to the infinity.

\begin{figure}[h]
	\centering
	\includegraphics[width=8.0cm]{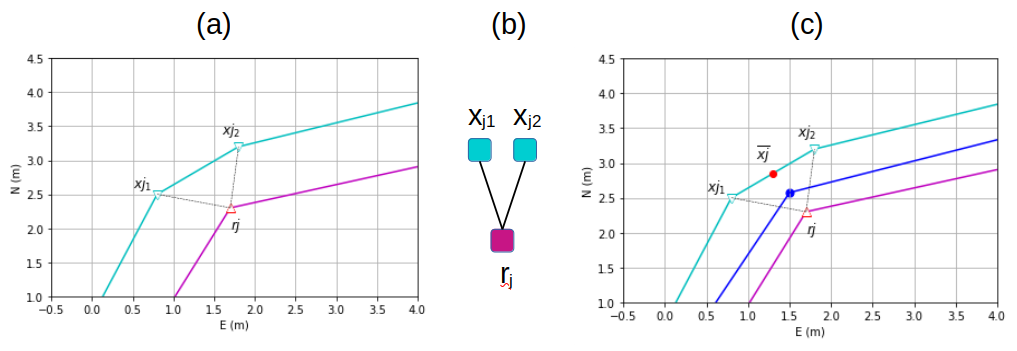}
	\caption{(a) A matching link between trajectory $\mathcal{X}$ and master trajectory $\mathcal{R}$. (b) A connected component of the bipartite graph associated with the matching. (c) representative and aggregation positions for a reference position of the master trajectory $\mathcal{R}$.}
	\label{compconnexe}
\end{figure}


\noindent A connected component of the bipartite graph ($\mathcal{X}$, $\mathcal{R}$, $\mu$) corresponds exactly to one or more matching links with $ni$ positions of $\mathcal{X}$ and $mj$ positions of $\mathcal{R}$. For example, in Figure \ref{compconnexe}-a and \ref{compconnexe}-b, the $j$-th connected component is composed of vertex ($rj_1$) from $\mathcal{R}$, vertices ($xi_1$ et $xi_2$) and the two associated vertex.

\subsection{Second step: representative selection on trajectory sections}

\noindent At the end of matching process, positions of trajectory $\mathcal{X}$ can be linked to many points of $\mathcal{R}$ and reciprocally. Thus, the second step of algorithm iteration is to choose for each connected component previously established, a representative position of each group of vertices from the trajectory $\mathcal{X}$: $\overline{xj}$. This ensures that the final aggregation is not too much influenced. If the segment of the trajectory is composed of several vertices, the representative is calculated from all the positions of the segment, otherwise the single position is chosen. \cite{etienne:hal-02110086}, take the center of gravity of the segment. In Figure \ref{compconnexe}-c, the red circle $\overline{xj}$ represents the center of gravity of the vertices $xi_1$ and $xi_2$). 

\subsection{Third step: aggregation of representative points}

\noindent At this point, all positions  of the master trajectory $(\mathbf{r}_j)_{j=1..m}$ have a unique homologous point $\overline{xj}$ from each trajectory $\mathcal{X}$. The positions of the new merged trajectory are calculated from the median of matching points. The median aggregation operator is better suited to handle outliers. \cite{etienne:hal-02110086} add a constraint in this step of the algorithm: each position aggregated must also be part of existing trajectory positions, so as not to be located in an unlikely place.

\vspace{0.5\baselineskip}

\noindent At last, we iterate using the aggregated trajectory as the new master trajectory. The algorithm \textbf{stops} when the distance between two subsequent estimation of the aggregated trajectory is below a predefined threshold. The distance used is the pointwise $L_2$ distance. 


%% file: 4_MIAA.tex
\newpage
\section{MIAA: a Modular and Iterative Aggregation Algorithm}
\label{MIAA}

\noindent Knowing the diversity of similarity and aggregation method for trajectories, we adapt the existing algorithm proposed in \cite{etienne:hal-02110086} by transforming it into a modular one. Modularity has substantial advantages such as flexibility (\textit{i.e.}a measure can be easily replaced by a another), scalability (\textit{i.e.} new measures can be easily added), and more relevant to the goal of this study, experiment testing (\textit{i.e.} study the behavior and the influence of measures and parameters in different contexts and with different data).

\begin{figure}[h!]
	\centering
	\includegraphics[width=8.2cm]{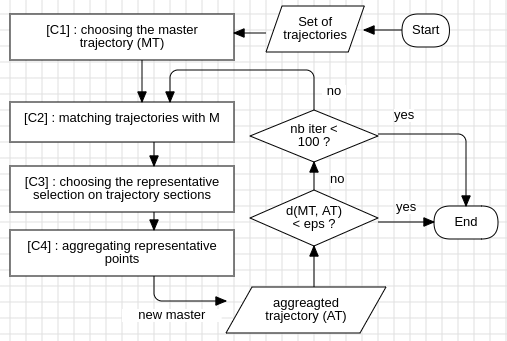}
	\caption{The four components of modular and iterative aggregation algorithm for GNSS trajectories.}
	\label{flow}
\end{figure}

\noindent Figure \ref{flow} illustrates the proposed Modular and Iterative  Aggregation Algorithm (MIAA) for GNSS trajectories which is composed by four components with new options. Thus, this proposal will create many variants of the algorithm. Note that, each component corresponds to a step of the algorithm.

\vspace{1\baselineskip}

\noindent The open-source implementation of the MIAA algorithm have been integrated into the Tracklib library (\cite{menerouxtracklibv2}).
 
\subsection{Component 1 - choosing the master trajectory} 

\begin{figure}[h!]
	\centering
	\includegraphics[width=7.5cm]{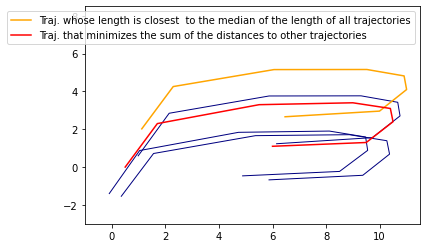}
	\caption{Heuristics for choosing a trajectory \textit{master}}
	\label{fig:c1}
\end{figure}

\noindent This component is representing the first step of the workflow. We propose three options (\textit{Cf.} Figure \ref{fig:c1}) to get the master trajectory. The first, already proposed in \cite{etienne:hal-02110086}, selects as master, the trajectory whose length is closest to the median of the lengths of all trajectories to be aggregated. We add, to this initial option, two new options: (1) the master trajectory is one that minimizes the sum of distances to other trajectories and (2) the master trajectory is randomly selected from the set of trajectories to be aggregated. These features have the advantages to study the influence of the choice of the master trajectory on the final result. 

\newpage
\subsection{Component 2 - matching trajectories with the master trajectory} 

\noindent This component is linked to the second step of the algorithms and contains different measures to compute the distance between two trajectories. Among the many similarity measures, we considered four measures: in addition to the matching based on the discrete Frechet distance, we add as new variants, two methods parameterized with $L_p$-norm ($p \in {1,2}$) and the nearest neighbour matching. 

\begin{figure}[h!]
	\centering
	\includegraphics[width=13cm]{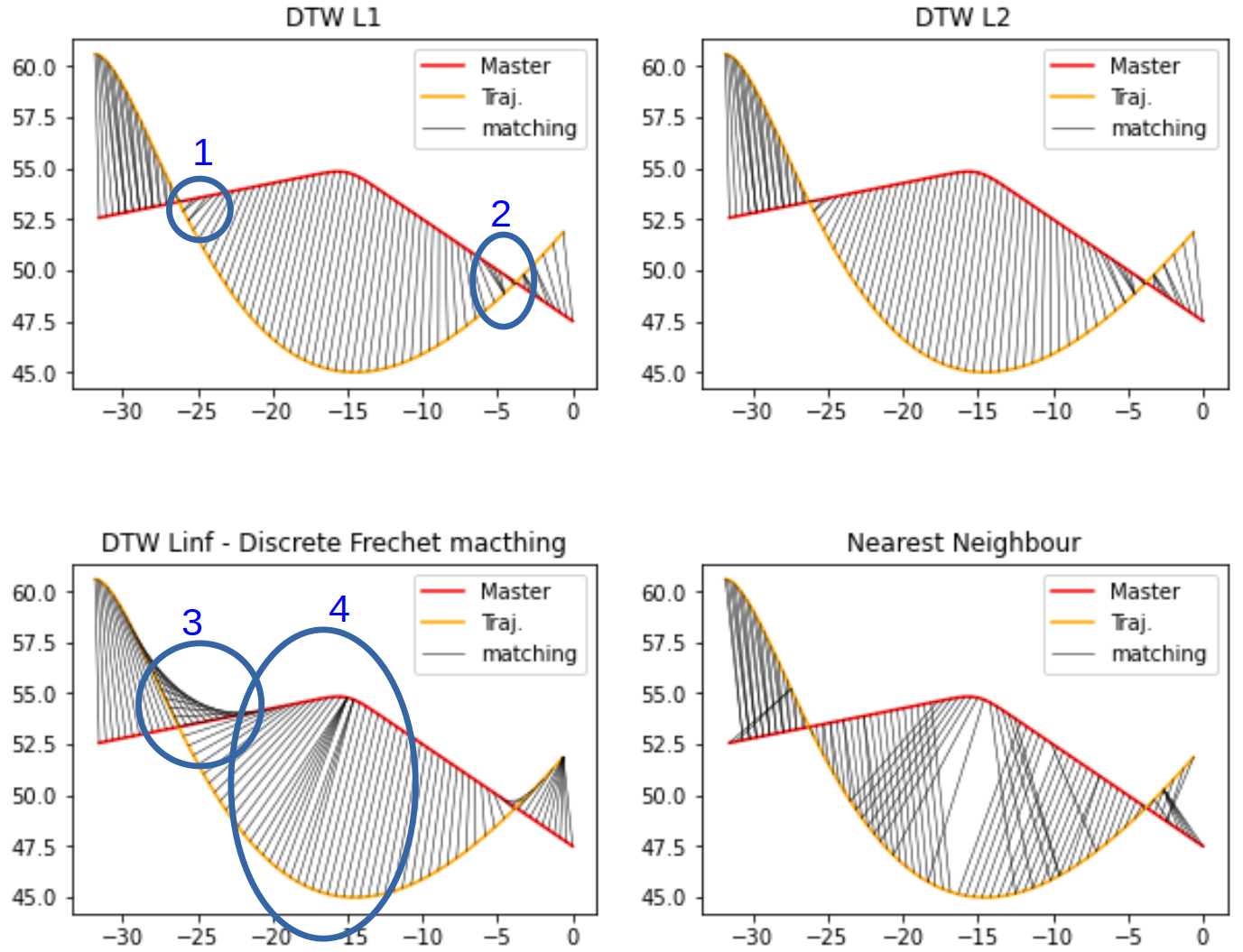}
	\caption{Heuristics for choosing a similarity measure for matching trajectories}
	\label{fig:C2}
\end{figure}

\vspace{1\baselineskip}

\noindent Clearly, the Nearest Neighbour distance differs from the others because it does not preserve the order of position in matching between the two trajectories, as described in the Figure \ref{fig:C2} (bottom right image). 

\vspace{1\baselineskip}

\noindent As mentioned in the section « Formalization of the original algorithm », 
Discrete Fréchet distance seeks to minimize the maximum of distances between the two curves. Once the maximum distance link is fixed, the other matching links are placed in the remaining spots. In Figure \ref{fig:C2}, the matching links are either grouped for distant points (Circled area 4) or create a large offset (Circled area 3).

\vspace{1\baselineskip}

\noindent DTW, both $L_1$ and $L_2$, aims at minimizing the sum of these distances,the matching links are therefore spaced more evenly, as shown in the the top 2 images in the Figure \ref{fig:C2}. The difference between the $L_1$ and $L_2$ distances in matching is minimal. The $L_2$ distance is the straight-line distance, while the $L_1$ distance follows the x-axis and y-axis. In Figure \ref{fig:C2}, the difference between these two distances becomes significant when the links between the points are close to the diagonals of the two axis (Circled area 1 and circle area 2).

\newpage
\subsection{Component 3 - choosing the representative position of each homologous points}

\noindent As mentioned in the second step of the iteration of the aggregation algorithm, it is possible that multiple points on a trajectory are matched to the same point in the reference trajectory. Then it is required to reduce these points to a single representative position. This can be done with different methods: center of gravity, position with median time, position furthest from the master trajectory.


\begin{figure}[h!]
	\centering
	\includegraphics[width=16cm]{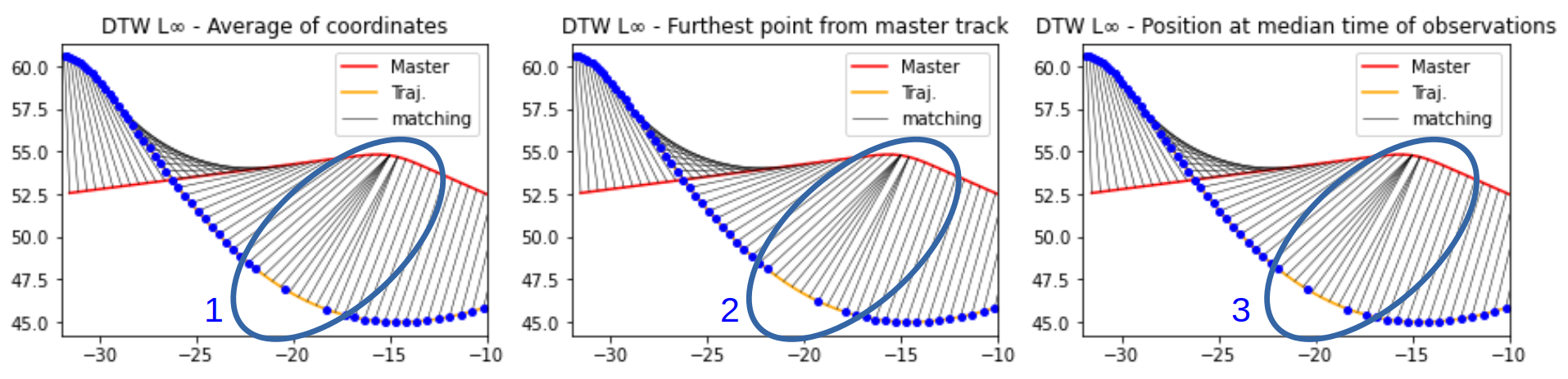}
	\caption{Heuristics for options in [C3] for DTW $L_\infty$}
	\label{fig:C3}
\end{figure}

\noindent With the Discrete Fréchet distance ($L_\infty$) in C2, Figure \ref{fig:C3} describe the three options. Unsurprisingly, the higher the link cardinality, the more this option will influence the position of the representative. The choices of the center of gravity and the position furthest from the master trajectory (Circled area 2) correspond to a spatial strategy, unlike the option of the position with median time. In Figure \ref{fig:C3}, since the points of the track are spaced at fixed time intervals, the difference between option 1 (Circled area 1) and option 3 (Circled area 3) is the same.

\subsection{Component 4 - aggregating the representative position}

\noindent So, for each reference position $(\mathbf{r}_j)_{j=1..m}$ on the master trajectory $\mathcal{R}$, we have a representative $\overline{xj}$ on each trajectory $\mathcal{R}$. Using an aggregation operator, a unique aggregating position can represent each reference position $\mathbf{r}_j$. We propose four aggregate function to calculate aggregating position: the marginal median (\cite{etienne:hal-02110086}), the geometric median, the $L_2$ mean and the $L_\infty$ which is the center of the minimum covering circle. In this component, we can add a constraint with the four approaches: whether or not to anchor the aggregated position to an existing position in the dataset's trajectories as explained in step 3 of \ref{LEAlgo}.

\begin{figure}[h!]
	\centering
	\includegraphics[width=16cm]{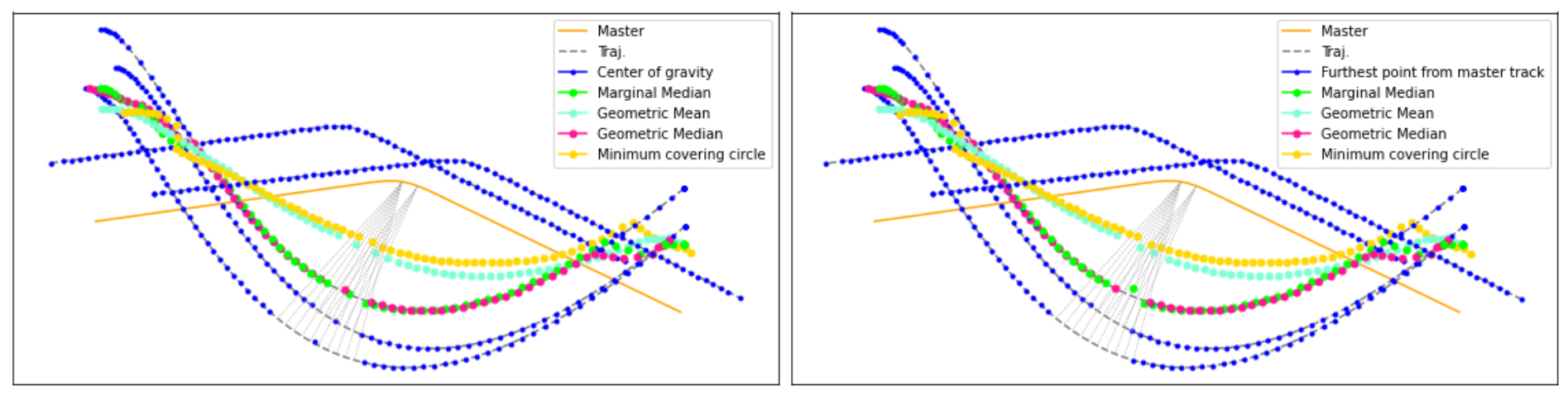}
	\caption{Heuristics for options in [C4] for DTW $L_\infty$}
	\label{fig:C4}
\end{figure}

\noindent This step is the one that determines the final position of the points in the track. In Figure \ref{fig:C4}, we can observe that the marginal median suggests a point among the positions of the representatives of C3, whereas the geometric mean proposes a new position for the trajectory, which can therefore be influenced by outliers.


%% file: 5_extension.tex
\newpage
\section{Extensive analysis and calibration}

\subsection{Termination of the algorithm}

\noindent It should be mentioned that, to our knowledge, the termination of the algorithm has not yet been studied in \cite{etienne:hal-02110086}. MIAA algorithm is an iterative algorithm, and there is no guarantee that it will converge from an IT point of view. For example, at the end of an iteration, you can return to a previous step (the merged trajectory corresponds to a previous master trajectory), therefore iterations are entering a cycle. We demonstrate this potential drawback with a counter-example described below.

\bigskip

\begin{figure}[h]
	\centering
	\includegraphics[width=12.5cm]{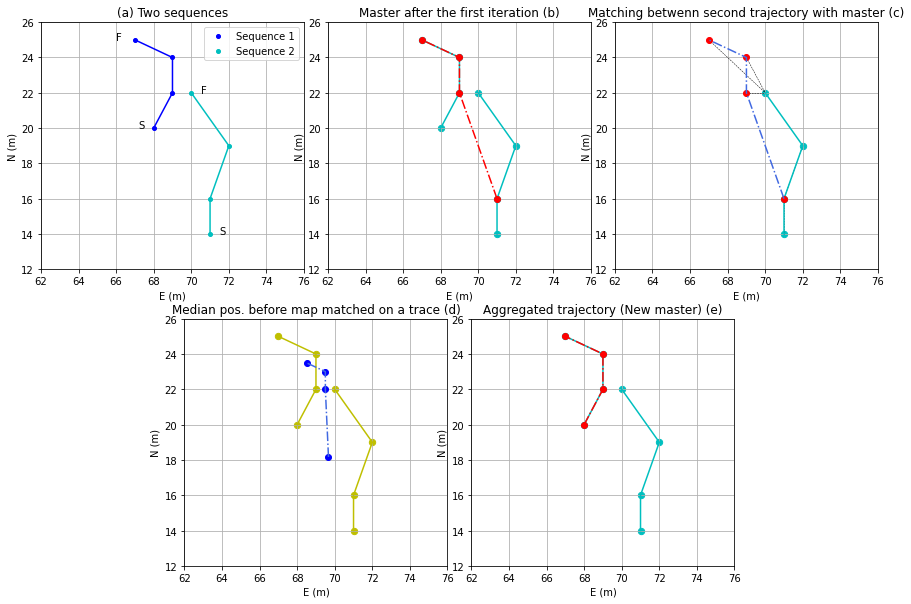}
	\caption{Counter-example showing a case of non-convergence}
	\label{conv1}
\end{figure}

\noindent For example, consider these two trajectories shown in Figure \ref{conv1}-a: 

\noindent \hspace*{4em} $\mathcal{X}$ = $<(68.0, 20.0), (69.0, 22.0), (69.0, 24.0), (67.0, 25.0)>$ \\
\hspace*{4em} $\mathcal{Y}$ = $<(71.0, 14.0), (71.0, 16.0), (72.0, 19.0), (70.0, 22.0)>$ 

\noindent The selected configuration is: for the master trajectory, is one that minimizes the sum of distances to other trajectories, the DTW-L1 has been chosen for matching trajectories, the center of gravity is taken for the representative position of each homologous and to aggregate the representative position, median aggregator is used to ultimately be map matched over a position of the trajectories. 

\smallskip

\noindent In this configuration, the first iteration of the MIAA algorithm returns the aggregated trajectory represented by the red line in Figure \ref{conv1}-b. Now let us take a closer look at the second iteration. The matching of the trajectories with the master trajectory produces a single matching link (one vertex of $\mathcal{X}_2$ with one vertex of $\mathcal{R}$) except for the first reference position, we have a matching link 1-3: ($r_1$, $x2_1$, $x2_2$, $x2_3$) (Figure \ref{conv1}-c) (corresponds to these three points of the second trajectory $x_1$ (71.0,14.0), $x2_2$ (71.0,16.0) and $x2_3$ (72.0,19.0)). Consequently the center of gravity of these three points is $\overline{x2_1}$ (214/3, 49/3) that is to say, the representative position associated to $r_1$ (Figure \ref{conv1}-d). 

\smallskip

\noindent With anchor constraint, we need to match the center of gravity $\overline{x2_1}$ with one of the four vertex: $r_1$, $x2_1$, $x2_2$, $x2_3$, respecting this order (IT implementation). There are two candidates positions with the same distance $r_1$ and $x2_3$ (6.1388...8), so the winner is $r_1$ (first in the list). 

\smallskip

\noindent At the end of the second iteration, the aggregated trajectory corresponds to the master trajectory of the first iteration. We are locked in a two-iteration cycle.

\subsection{Algorithm calibration: further details}

\noindent This section reproduces the online notebook on the Tracklib documentation website "Aggregated Trajectory: position errors and shape deviation estimation". It is inspired by the work presented in \cite{10.1145/3678717.3691325} and also provides some pieces of \textit{Python} code from Tracklib library (\cite{menerouxtracklibv2}) documentation\footnote{https://tracklib.readthedocs.io/en/latest/usecase/AggregatedTrajectory.html}. 

\medskip

\noindent We will examine the aggregation of trajectories using two matching distances: the Fréchet distance and the DTW-L2 distance, and thus see its ability to reconstruct accurately the common path followed by all the individual sample trajectories by comparing the position errors and the shape deviation.

\medskip

\noindent Our experiment will be conducted in four steps:
\begin{enumerate}
	\item Step 1: creating a synthetically reference track considered as the ground truth track,
	\item Step 2: creating a set of simulated tracks from the reference track, 
	\item Step 3: computing the aggregation track from the set previously constructed (step 2),
	\item Step 4: the error between the estimated and ground truth track is then evaluated.
\end{enumerate}

\subsubsection{Step 1: Create reference tracks}

\noindent For this experimentation, we will examine the trajectory aggregation process for different types of paths to be reconstructed. In the context of mountain hiking, we identified three characteristic path shapes: nearly straight segments, moderately sinuosity road segments, and a zigzagging path composed of a series of switchbacks. To maintain shape consistency throughout the entire road segment, the generated trajectories have a length of approximately 300 meters.

\medskip

\noindent This three commonly mountain path shapes correspond to the reference tracks.

\begin{python}
tkl.seed(123)
# ----------------------------------------------------------
# Generate the path 'Almost straight'
sentier1 = tkl.generate(0.5, dt=10)[::3]
sentier1.scale(5)

# ----------------------------------------------------------
# Generate the path 'Moderate sinuosity'
sentier2 = tkl.generate(0.1, dt=10)[::3]

# ----------------------------------------------------------
# Generate the path 'Switchbacks'
base_lacets = tkl.generate(0.4, dt=10)
sentier3 = tkl.noise(base_lacets, 20, tkl.SincKernel(20),
direction=tkl.MODE_DIRECTION_ORTHO)[::3]
sentier3.scale(4)

# ----------------------------------------------------------
SHAPES = ['Almost straight', 'Moderate sinuosity', 'Switchbacks']
sentiers = [sentier1, sentier2, sentier3]
\end{python}

\begin{figure}[h]
	\centering
	\includegraphics[width=10.0cm]{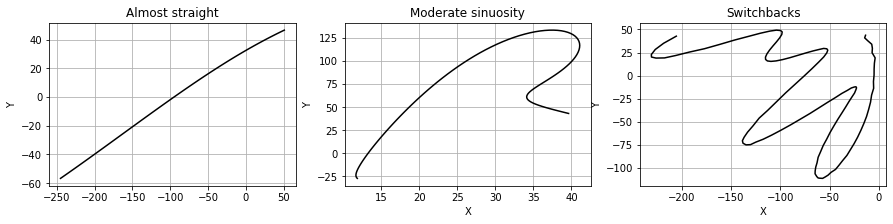}
\end{figure}

\subsubsection{Step 2: Create sets of simulated GNSS trajectories}

\noindent Now that we have our three reference trajectories, we want to create three sets of N simulated GNSS trajectories from them. To generate realistic noise, we used an approach described in \cite{ripley2009stochastic} and also employed in \cite{Meneroux2023}: 

\smallskip

\noindent with a random generator, we sampled $N$ i.i.d. unit-variance and zero-mean gaussian values, compiled in a vector $\mathbf{x}$. It can easily be shown that, for any positive-definite matrix $\mathbf{\Sigma} \in \mathbb{R}^{n \times n}$, the random vector $\mathbf{y} = \mathbf{Ax}$ where $\mathbf{A}$ is a Cholesky factor of $\mathbf{\Sigma}$, is a realization of a correlated random vector $\mathbf{Y}$ having covariance matrix $\Sigma$. The covariance matrix $\mathbf{\Sigma}$ is formed with a (stationary) covariance kernel with three parameters:

\begin{itemize}
	\item The \textbf{type of kernel}: exponential, gaussian, and triangular models are used.
	\item The \textbf{amplitude} of noise: is between 0 and 5 meters, as it is quite uncommon to find building databases with more than 5 m error amplitude. If necessary, the output tables could be extended to handle large errors.
	\item The correlation \textbf{scope} of the noise which roughly speaking describes how far apart two errors would remain correlated (in both amplitude and direction): between 1 m (white noise) and 1000 m (global translation).
\end{itemize}

\noindent In tracklib, you have to create a Kernel and use the noise method on a track.

\bigskip

\noindent Trajectories have been generated with a 5 m-amplitude exponential covariance process, completed by a 1 m white noise process, a 50 cm range Gaussian Process for referencement error.

\begin{python}
N = 20

# Generate 'Almost straight'
tracks1 = tkl.core.TrackCollection([sentier1]*N)
tracks1.noise(5, tkl.ExponentialKernel(1))
tracks1.noise(1, tkl.GaussianKernel(0.5))

# Generate 'Moderate sinuosity'
tracks2 = tkl.core.TrackCollection([sentier2]*N)
tracks2.noise(5, tkl.ExponentialKernel(1))
tracks2.noise(1, tkl.GaussianKernel(0.5))

# Generate 'Switchbacks'
tracks3 = tkl.core.TrackCollection([sentier3]*N)
tracks3.noise(5, tkl.ExponentialKernel(1))
tracks3.noise(1, tkl.GaussianKernel(0.5))
\end{python}

\begin{figure}[h]
	\centering
	\includegraphics[width=12.0cm]{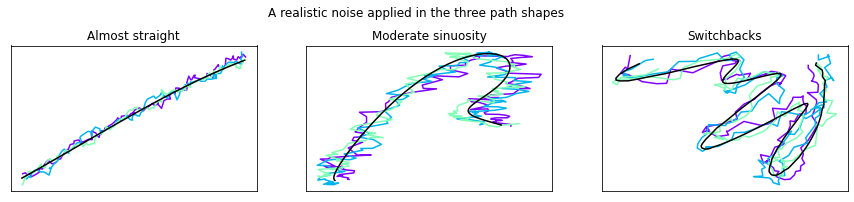}
\end{figure}

\subsubsection{Step 3: Compute aggregated trajectory}

\noindent We have two evaluations to do: compute the aggregated trajectory with Frechet distance and with $L_2$ distance. 

\bigskip

\noindent Each one is conducted as follows: we generate $N'$ random noisy traces, with $N' \in [1, E]$ and $E \le N$. For each generation, we are going to compare the agregated trajectory against the reference track (*i.e.* the ground truth). 

\bigskip

\noindent This code compute an aggregation track for a set of 10 trajectories: 

\bigskip

\begin{python}
FRECHET = {'Almost straight': tkl.TrackCollection(),
	'Moderate sinuosity': tkl.TrackCollection(), 'Switchbacks': tkl.TrackCollection()}
DTW = {'Almost straight': tkl.TrackCollection(),
	'Moderate sinuosity': tkl.TrackCollection(), 'Switchbacks': tkl.TrackCollection()}

represent_method = tkl.MODE_REP_BARYCENTRE
aggmeth = tkl.MODE_AGG_MEDIAN
cstt = False
master = tkl.MODE_MASTER_MEDIAN_LEN
itermax = 25

# create set of 10 trajectories
TAB = set()
while len(TAB) <= 10:
    n = randint(0, N-1)
    TAB.add(n)

sets1 = tkl.TrackCollection()
for idx in TAB:
    sets1.addTrack(tracks1[idx])

# Frechet
p = float('inf')
mode = tkl.MODE_MATCHING_FRECHET
central1 = tkl.fusion(sets1, master=master, dim=2, mode=mode, p=p,
        represent_method=represent_method, agg_method=aggmeth, constraint=cstt)
FRECHET['Almost straight'].addTrack(central1)
...

# DTW
p = 2
mode = tkl.MODE_MATCHING_DTW
central1 = tkl.fusion(sets1, master=master, dim=2, mode=mode, p=p,
	    represent_method=represent_method,  agg_method=aggmeth, constraint=cstt)
DTW['Almost straight'].addTrack(central1)
...
\end{python}

\begin{figure}[h]
	\centering
	\includegraphics[width=10.0cm]{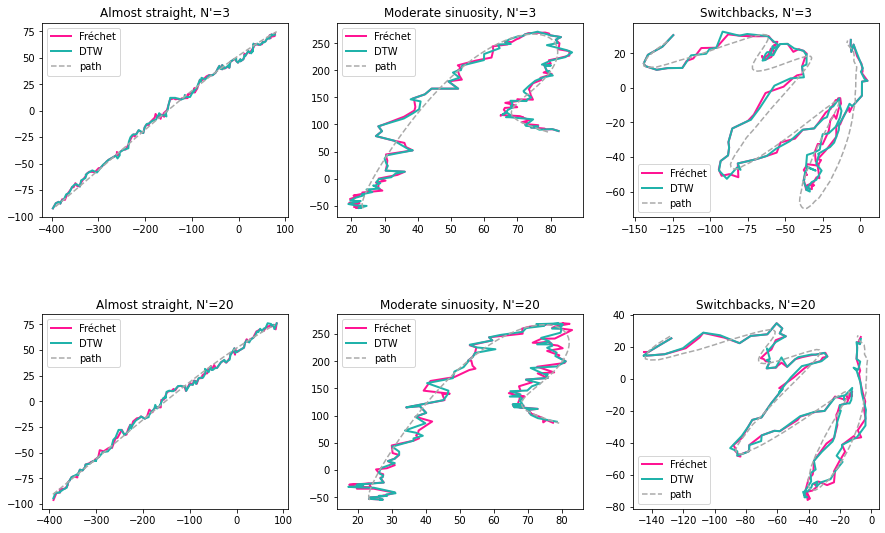}
	\caption{Aggregated trajectories for two samples of 3 and 20 trajectories and computed with Discrete Fréchet distance (in pink) and DTW-$L_2$ distance (in blue)}
\end{figure}


\newpage
\subsubsection{Step 4: Evaluate error between the aggregated and ground truth track}

\smallskip

\noindent \underline{\textbf{Position error measurement}}

\bigskip

\noindent We compute the distance $pointwise L_2$ (\textit{Root Mean Square Error}) by finely resampling both the aggregated and ground truth trajectories from 120 to 1,000 points, ensuring one point every 50 cm over 500 meters.

\bigskip

\begin{python}
def rmse(central, sentier):

    central.resample(npts=1000, mode=1)
    sentier.resample(npts=1000, mode=1)

    # compute the distance NearestNeighbour
    m = min(sentier.size(), central.size())
    return tkl.compare(central[0:m], sentier[0:m], tkl.MODE_COMPARISON_POINTWISE, p=2)

rmse(FRECHET['Almost straight'][s], sentier1)
...
rmse(DTW['Switchbacks'][s], sentier3)
\end{python}

\bigskip

\noindent Which results in the following graphically:

\begin{figure}[h]
	\centering
	\includegraphics[width=15.0cm]{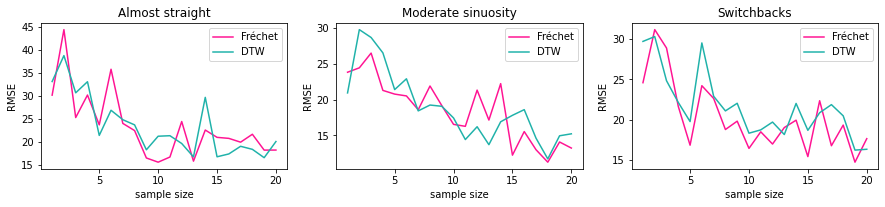}
	\caption{Position error measurement for each samples of $N$ trajectories generated}
\end{figure}

\noindent From the perspective of positional accuracy, neither matching with the Fréchet distance nor matching with the DTW distance prevails: the two curves have the same shape, and the differences, regardless of the number of trajectories in the sample, are minimal.

\newpage
\noindent \underline{\textbf{Shape deviation measurement}}

\bigskip

\noindent Let's start by aligning, with a geometric affine transformation, the aggregated track with the reference track to abstract away from position issues. Then, we finely resample both the aggregated and ground truth trajectories from 120 to 1,000 points, ensuring one point every 50 cm over 500 meters. Finally, we compute the distance *NearestNeighbour* for all positions to estimate shape deviation.

\bigskip

\begin{python}
def shapeDeviationMeasure(central, sentier):
    
    # Align the aggregated track with the reference track
    tkl.mapping.mapOn(central, sentier, verbose=False)

    # resample to 1000 points
    central.resample(npts=1000, mode=1)
    sentier.resample(npts=1000, mode=1)

    # compute the distance NearestNeighbour
    return tkl.compare(central, sentier, tkl.MODE_COMPARISON_NN, p=2)

shapeDeviationMeasure(FRECHET['Almost straight'][s], sentier1)
...
shapeDeviationMeasure(DTW['Switchbacks'][s], sentier3)
\end{python}

\bigskip

\noindent Which results in the following graphically:

\begin{figure}[h]
	\centering
	\includegraphics[width=15.0cm]{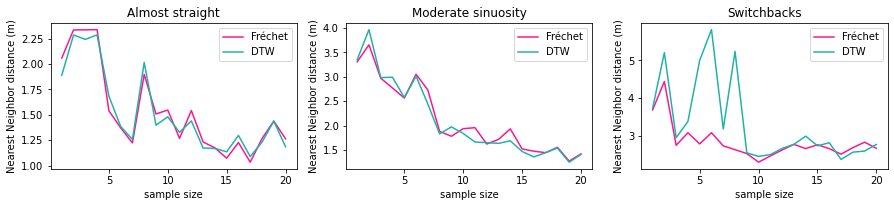}
	\caption{Shape deviation measurement for each samples of $N$ trajectories generated}
	\label{fig:shape}
\end{figure}

\noindent The Figure \ref{fig:shape} still shows a close similarity between the two curves: shape deviation measurement are quite close together and their profiles are quite similar.

\newpage
\subsection{Crowd-sourced trajectories dataset} 

\begin{figure}[!htb]
	\begin{tabular}{cc}
		\begin{minipage}{0.48\textwidth}
			\begin{subfigure}{\linewidth}
				\centering
				\includegraphics[width=7.5cm]{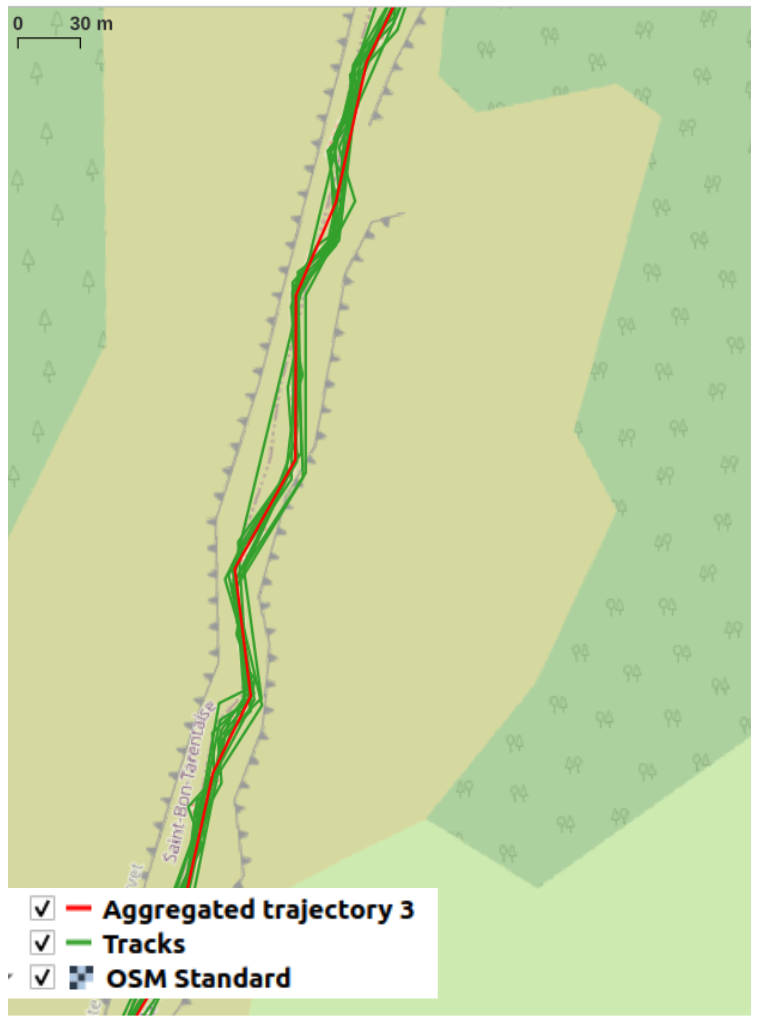}
				\caption{Aggregated trajectory closely follows the ridge path}
				\label{fig:vgi1}
			\end{subfigure}
		\end{minipage}
		& 
		\begin{minipage}{0.48\textwidth} 
			\begin{subfigure}{\linewidth}
				\centering
				\includegraphics[width=0.9\textwidth]{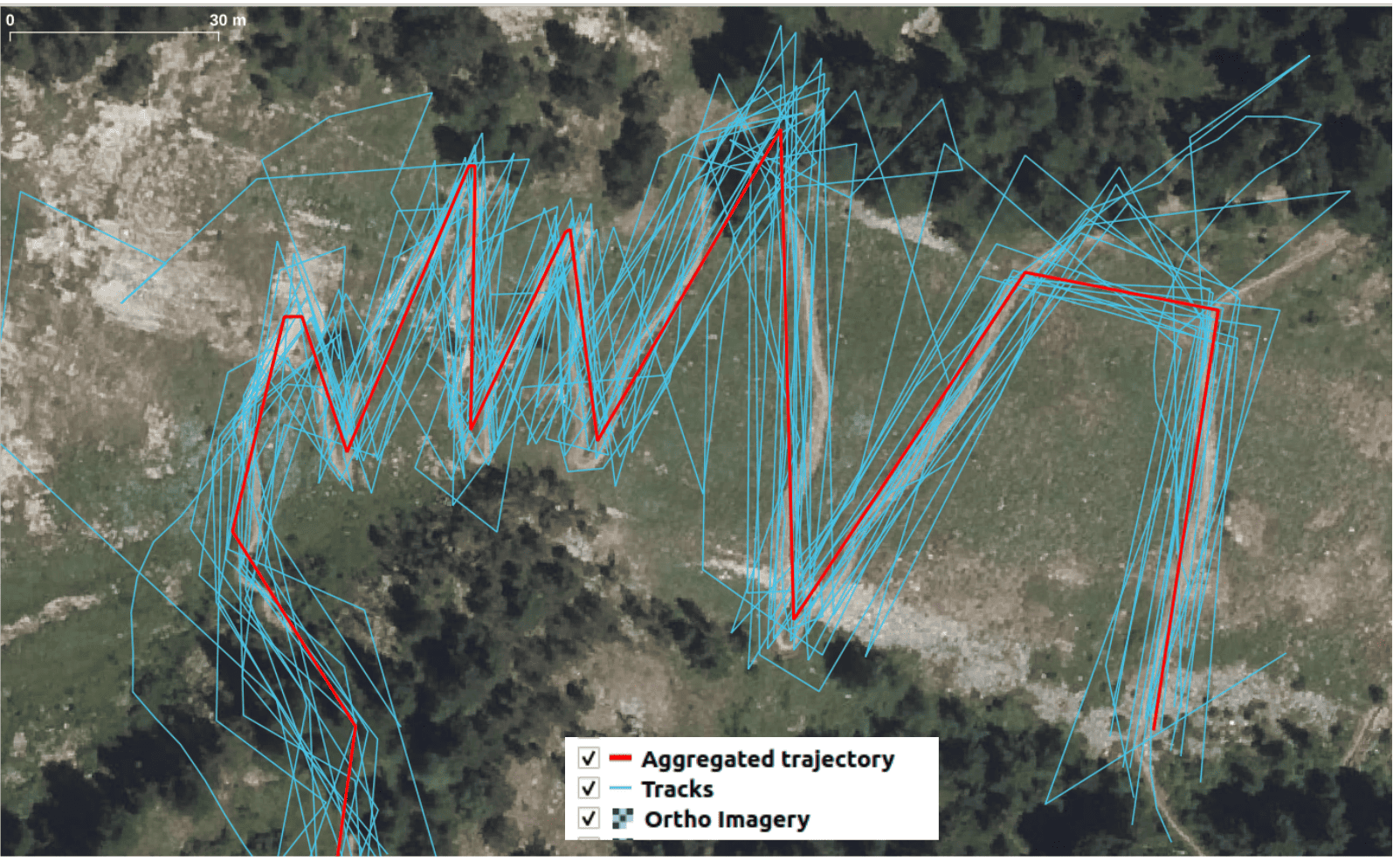} 
				\caption{Aggregated trajectory in red follow the most popular route}
				\label{fig:vgi2}
			\end{subfigure}
			\vspace*{0.6cm}
			\begin{subfigure}{\linewidth}
				\centering 
				\includegraphics[width=0.9\textwidth]{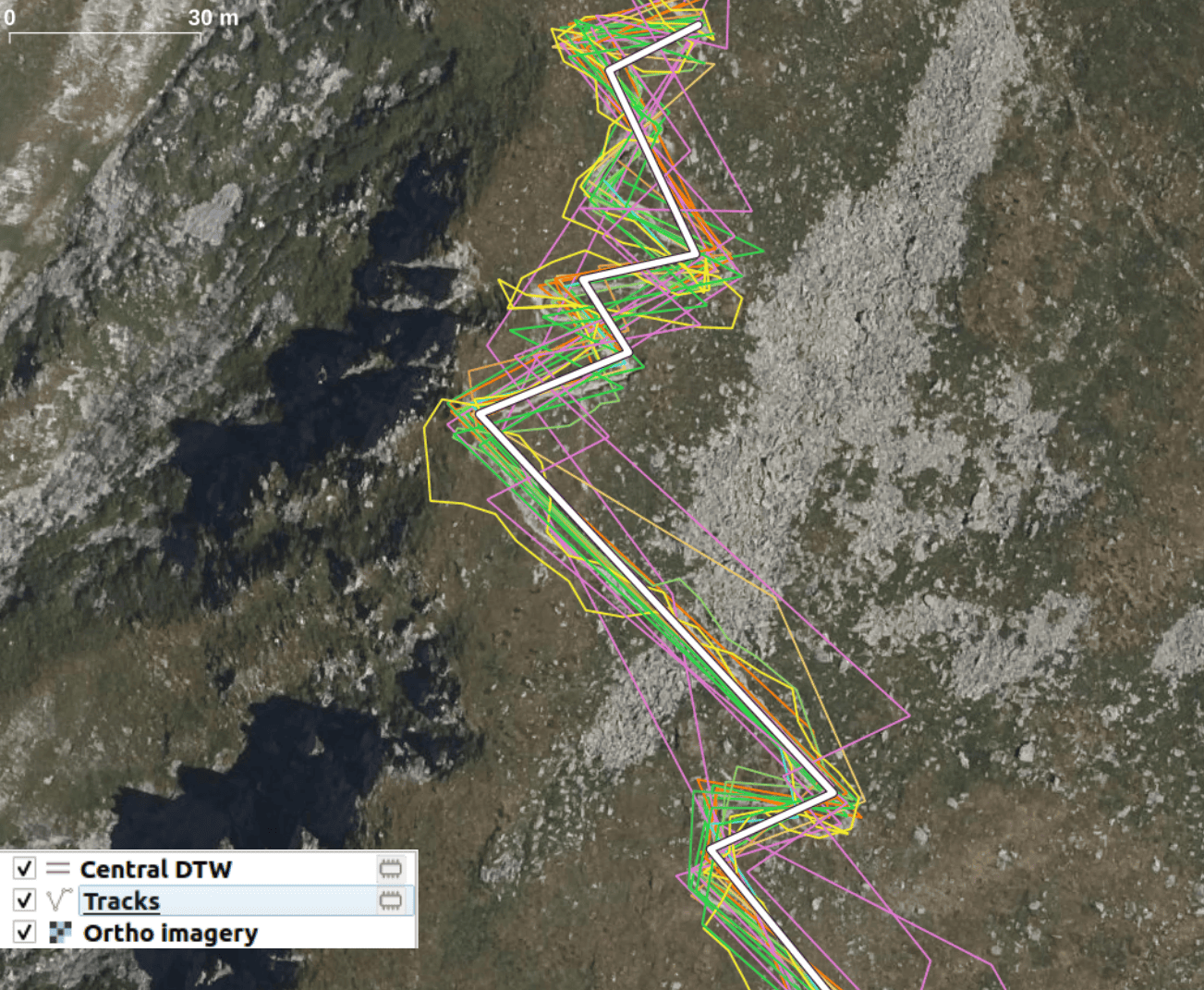}
				\caption{median operation effect in a series of bends with a Crowdsourced dataset}
				\label{fig:vgi3}
			\end{subfigure}
		\end{minipage}
	\end{tabular}
	\caption{Aggregated trajectory result on Crowd-sourced dataset}
\end{figure}

\noindent Here we present the results obtained on real data. As already mentioned, the analyse is not strictly "metrological" but rather qualitative, a graphic investigation. We define a first context: a ridge which characterises terrain constraints (Figure \ref{fig:vgi1}), a switchback trail which characterises an infrastructure support (Figure \ref{fig:vgi2}) and a straight line followed by a series of twists and turns with varying distances between them which characterises variations of the shape of the path (Figure \ref{fig:vgi3}).

\vspace{0.5\baselineskip}

\noindent Note that, traces may have been previously filtered via a simplification algorithm like Douglas-Peucker. Indeed, all geometries offer the same characteristics like vertex of bends, turns in route. If this was not the case, traces would be more dissimilar in shape.

\noindent Aggregated trajectory produced by the MIAA algorithm (red line in Figure \ref{fig:vgi2} and white line in Figure \ref{fig:vgi3}) detects the road shape very well, all the bends have been detected. In the context of mountain hiking, we can see the aggregated trajectory deviates from the road, which may make it possible to quantify the off-tracking. In Figure \ref{fig:vgi1}, we can observe that some turns are not detected, as the geometry of the aggregated trajectory tends to generalize and lose certain inflection points.

\newpage
\subsection{Perspective: performance under extreme conditions} 

\noindent Concerning high noise amplitude, we deliberately chose a low one (5 meters) in our experiments to adapt to advancements in GNSS technological (new constellations, new signals, improvements in receiver electronics). However, we have not studied the case of extreme noise conditions, and we could relaunch computations with extreme values, on the same synthetic GNSS trajectories datasets, with an amplitude of the exponential covariance process from 5 meters (\textit{Cf.} \cite{journalpone0283090}) to 30 meters (as mentioned in the literature for errors in heavy coverage canopy areas in \cite{Piedallu2005}). 

\vspace{1\baselineskip}

\noindent Additionally, for outlier traces, particularly off-roads, we conducted empirical tests on the multi-sensors/multi-canopy traces dataset. With few outliers, our tests confirmed there was no impact on the aggregated trajectory, demonstrating its robustness as mentioned in \cite{etienne:hal-02110086}. The algorithm computes, for each matched points in the master trajectory, the robust median positions between all trajectories. Thus, we decided not to further investigate robustness, in order to focus on its modularity. We could study the performance of the algorithm with many off-roads in the sample traces to be merged. For example, the robustness analysis would consist in including a proportion p of off-roads in a given number of trajectories, and determine how the quality of the aggregated trajectory decreases as p increases. In our opinion, with the effect of the median, results should not be dramatically modified, at least up to p=25\%. If needed, we can run the test and add comments with a figure depicting the aggregated trajectory quality as a function of p.